\documentstyle[11pt,paspconf]{article}

\begin{document}

\title{Electron Capture in Early Gravitational Collapse --
Nuclear Equation of State}

\author{A. Ray \altaffilmark{1}}
\affil{NASA/Goddard Space Flight Center, Greenbelt, MD 20771} 

\author{F. K. Sutaria and J. Sheikh}
\affil{Tata Institute of Fundamental Research, Bombay 400005, India}

\altaffiltext{1}{NAS/NRC Senior Research Associate, on leave of absence
from Tata Institute of Fundamental Research, Mumbai (Bombay)} 


\begin{abstract}
We present the spectra of pre trapping neutrinos
emitted from a core collapse supernova (having main
sequence masses 15 and 25 $M_{\sun}$) within 1 kpc
which can be detected by terrestrial detectors. 
The neutrino spectrum depends on the abundance of nuclei and free
protons which undergo electron capture which in turn is determined
by nuclear properties of the stellar core.
The ambient temperature in the early pre trapping phase
is not so high as
to wipe out shell and pairing effects. We present results from Relativistic
Mean Field (RMF) calculations, which we use to predict properties of the neutron
rich nuclei which dominate the stellar composition at this stage of stellar
collapse and compare the RMF results with 
the Baron et al (BCK) equation of state. 

\end{abstract}


\keywords{supernovae, neutrinos, equation of state}


\section{Detectable number of neutrinos and their spectra}
 Neutrinos emitted from the early phase of core collapse that
precedes a type II/Ib/Ic SN explosion, 
are emitted mainly due to e$^-$-captures on free 
protons and heavy nuclei (in the f-p shell with $A \geq 60$).
Up to core densities
of $\simeq$ 3 $\times 10^{11}$ gm/cc
(neutrino trapping density) these $\nu_e$ escape
freely from the overlying
stellar matter without any interaction that change their
energy. 
Their spectroscopy by terrestrial detectors would yield
important information on the physical
and the nuclear configuration of the collapsing
stellar core. 

The total number of neutrinos emitted from a 1.4 $M_{\odot}$ stellar
core as it evolves from a initial density of $\sim 4 \times 10^9 \rm g/cm^3$ 
to a  neutrino-trapping density of $\simeq 2 \times 10^{11} \rm g/cm^3$
is $\simeq 5 \times 10^{55}$. 
The charge-current 
reaction $\nu_e(d,pp)e^-$ on deuterium
nuclei in the Sudbury Neutrino Observatory (SNO)  D$_2$O detector  
and the $\nu_e -e$ scattering reaction for the more massive Super
Kamioka (H$_2$O based) detector can facilitate the detection of a significant
number of these neutrinos at 1 kpc (Burrows 1990). 

The neutrino production rate (i.e. the spectrum of the emitted $\nu_e$)
as a function of neutrino energy $E_{\nu_e}$ depends on the electron capture 
rate for heavy
nuclei ($H$) and free protons ($fp$) which are given by 
\begin{equation}
\lambda_{fp, H} = { \log 2  \over (ft)_{fp,H} } {<G> \over (m_ec^2)^5}
{ E_{\nu}^2 (E_{\nu} + Q_{fp,H}) \sqrt{ (E_{\nu} + Q_{fp,H})^2 - (m_e
c^2)^2) } \over (1 + \exp( E_{\nu} + Q_{fp,H} - \mu_e)) } d E_{\nu}
\label{eq: rateHp}
\end{equation}
where $<G>$ is the coulomb correction factor (taken to be $\simeq 2$ for heavy
nuclei and $=1$ for free protons, Fuller, 1982) and the ft's are related
to the nuclear transition matrix element. Fig 1(a) presents  the 
incident cumulative
spectra of $\nu_e$ emitted up to a density of $2.4 \times 10^{11} \rm g/cm^3$, 
from a supernova explosion 1 kpc away. These calculations are done using
a single zone approximation for the collapsing stellar core 
(Ray et al., 1984), and approximating the 
ensemble of neutron-rich heavy nuclei present in the 
core by a mean nucleus A (which is the most abundant nucleus
present under the given thermodynamic core configuration -- Bethe et al. 1979). 
The spectra for 
free proton has been weighted by the relative fraction of free protons 
$X_p$ present in the core.
$Q_{fp,H}$ is the e$^-$- capture Q-value for free-protons and heavy nuclei
respectively  and Q$_H$ is given as :
$ Q_H = \hat \mu + 1.297 + E_{GT}$
where $\hat \mu$ ($=\mu_n -\mu_p$)  is the difference in the neutron and
proton chemical potentials
and $E_{GT}$ is the energy of the Gamow-Teller Giant Resonance centroid
(the centroids in fp-shell nuclei, found from experimental data
from (n,p) reactions have been used for characterizing GT centroids
in fp-shell nuclei (Sutaria and Ray, 1995) and are close to 3 MeV as used here).
For free protons this reduces to $Q_{fp} = \hat \mu + 1.297$.

\begin{table}
\caption{Pre-trapping
neutrino detections in SNO and Super Kamioka with hardness ratios
up to $\rho_{10}$ = 24.16 for indicated heavy nuclear e-capture
matrix elements for 15 M$_{\odot}$ Fuller (1982) and 25 M$_{\odot}$
Weaver et al (1985) pre supernova stars.}
\vskip 1 true cm
\begin{center} \scriptsize
\begin{tabular}{c|c|c|cc|cc|cc}
Star Mass  &$|M_{GT}|^2  $ &$t_{collapse}   $ &
\multicolumn{2}{|c|} {Pre-trapping Variables}&
\multicolumn{2}{|c|} {No. Detected} &
\multicolumn{2}{|c|}{Hardness Ratio\tablenotemark{a}} \\
& & (ms) & $Y_{ef}$& $S_f/k_B$ &
SNO & S-K & SNO & S-K \\
15 M$_{\odot}$ & 1.2/0.1 &120 & 0.3909   & 1.0021   & 82     &394       &0.2786
&0.8540  \\
               & 2.5/0.1 &120 & 0.3896   & 1.0085   & 66     &344       &0.2876
& 0.9537  \\
25 M$_{\odot}$ & 1.2/0.1 &190 & 0.3828   & 1.1080   &120     &566       &0.2878
&0.8319  \\
               & 2.5/0.1 &190 & 0.3813   & 1.1204   & 99     &499       &0.2916
&0.9190 \\
\end{tabular}
\end{center}

\tablenotetext{a}{The hardness ratio denotes the number
of neutrino events in the 5 MeV $\leq E_{\nu_e} \leq 12$ MeV
and 12 MeV $\leq E_{\nu_e} \leq$ 25 MeV bands.}
\end{table}
Fig.1(b) and Fig.1(c) show the expected $\nu_e$ spectra in SNO 
and Super Kamioka detectors respectively and are obtained by folding the 
spectra in Fig 1(a) with the energy dependent detection cross-sections
quoted in Burrows (1990) and Sehgal (1974). 
Table 1 displays the total number of neutrinos which can be expected 
to be detected in the SNO and Super-Kamioka detectors. The
neutrino production rate depends on the e$^-$-capture rate, and hence on
the weak interaction strength ($\propto |M_{GT}|^2$) of the nuclei undergoing 
$e^-$-capture. The numbers in Table 1 have been calculated for two
different values of $|M_{GT}|^2$. Nuclei beyond $^{74}$Ge become  neutron 
shell-blocked against 
e$^-$-capture (Fuller 1982), and 
since capture can proceed only via
either thermally
excited shell unblocking or via forbidden transitions, the $e^-$-capture
strength drops to $|M_{GT}|^2 =0.1$ for neutron shell
blocked nuclei (Kar and Ray 1983).
\begin{figure}
\vspace{12.7 cm}
\caption{Neutrino spectra: Incident, in SNO and in Super-Kamioka.} 
\caption{ Comparison of $\mu_n$ and $\hat \mu= \mu_n-\mu_p$ 
using RMF and BCK equation of state at $\rho= 1.0 \times 10^{10} 
\rm g/cm^3$.}
\end{figure}

\section{RMF calculations of core nuclear properties}
The neutrino
spectrum depends on the abundances of nuclei undergoing
electron capture, and this in turn is determined by the  nuclear properties
specifically, the electron capture Q-values $Q_{fp,H}$ and the 
nuclear chemical potentials $\mu_n$ and $\hat \mu$. At the stage of collapse 
considered, the ambient temperature $T$ is typically less than 
about 1 MeV, and in this range of temperatures, the nuclei undergo transition
from a region where shell and pairing affects dominate to the region where the
nucleus can be well approximated by the liquid drop model. To take into account the 
shell and pairing effects in the early 
stage of collapse, we have used the Relativistic Mean Field Theory 
(see e.g., Sheikh at al. 1993) to calculate $\mu_n$ and $\hat \mu$
for a number of nuclei ranging from $^{52}$Mn to $^{74}$Ge 
(Sutaria et al. 1997). 

In Fig. 2(a) we present the
calculation of $\mu_n$ for isotopes of Ni in the ground state
which are expected to have significant
abundance in this early stage of stellar collapse. Fig. 2(b) presents the 
values of 
$\hat \mu$ for the same set of nuclei. 
Earlier workers (Baron et al. 1985, Cooperstein, 1985) had developed nuclear 
equations of 
state which were based on the liquid drop model of the nucleus, with corrections
for  nuclear compressibility etc. In these models, the nuclear 
energy per nucleon $W_N$ is a function of the nuclear density $\rho_N$, the 
nuclear volume $V_N$, the proton fraction $x=Z/A$  and the ratio 
$u=\rho/\rho_N$ corresponding to stellar packing fraction.
The chemical potentials of the neutrons and protons in equilibrium with
nuclei under given thermodynamic conditions have been calculated in BCK
code by 
the usual methods (see e.g. Bethe et al., 1979 and references therein).
A comparison of the $\mu_n$ and $\hat \mu$ from the
BCK equation of state with the RMF computations is 
displayed in figures 2(a) and 2(b).  
In these figures, we have used outputs of BCK code 
at low temperature limit ($T \sim 0.1$ MeV)
corresponding to the nuclear parameters:
$a_V =-16.0$, S$_V$ (volume symmetry energy) = 30.34, and the nuclear
compressibility factor K$_0$ = 180.0 MeV.
Properties of Ni and other neutron rich nuclei in the fp-shell, such as
binding energies, nuclear deformations etc, computed from the RMF
code were discussed in relation to phenomenological nuclear models
by Sutaria et al (1997).

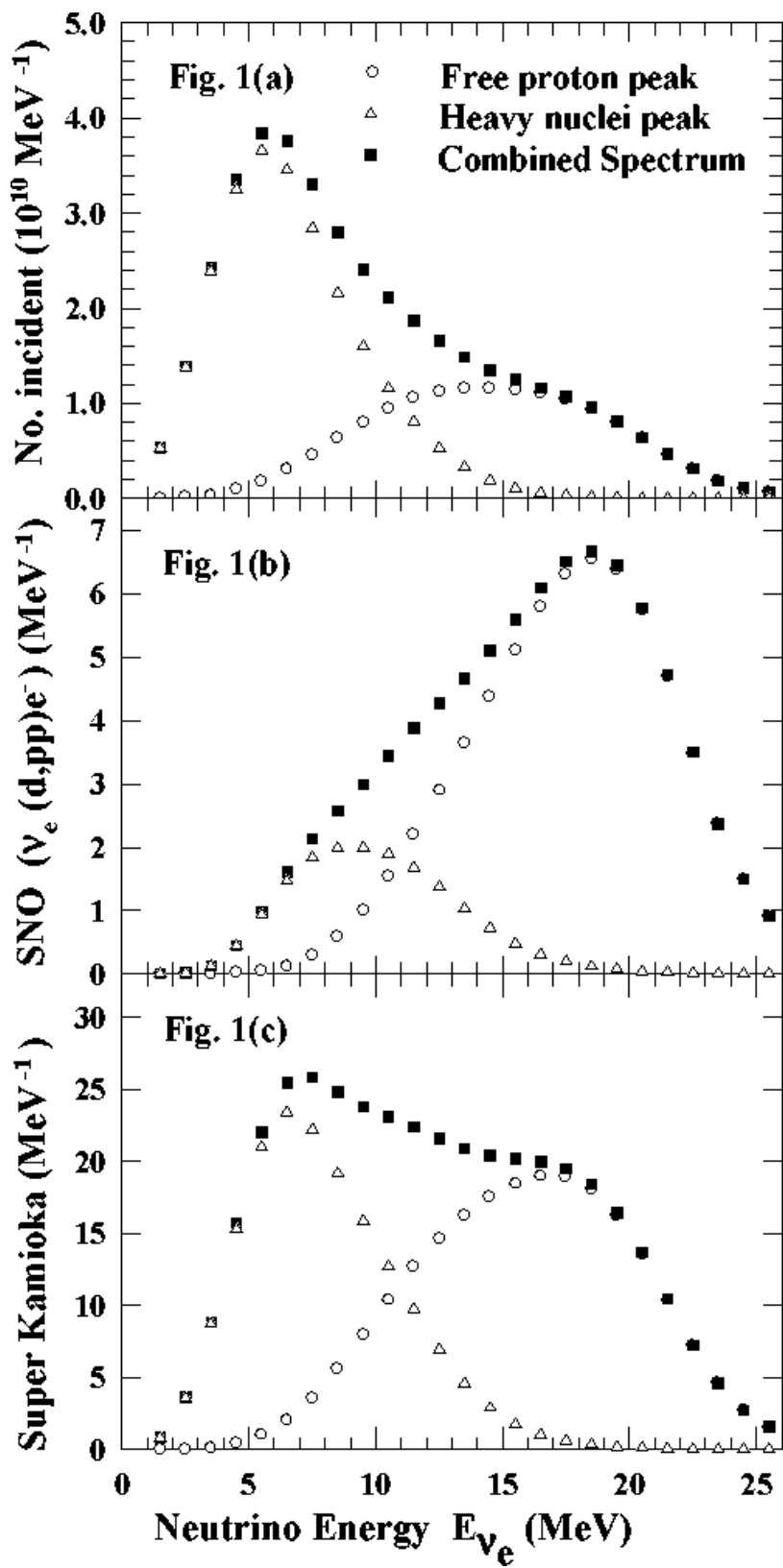
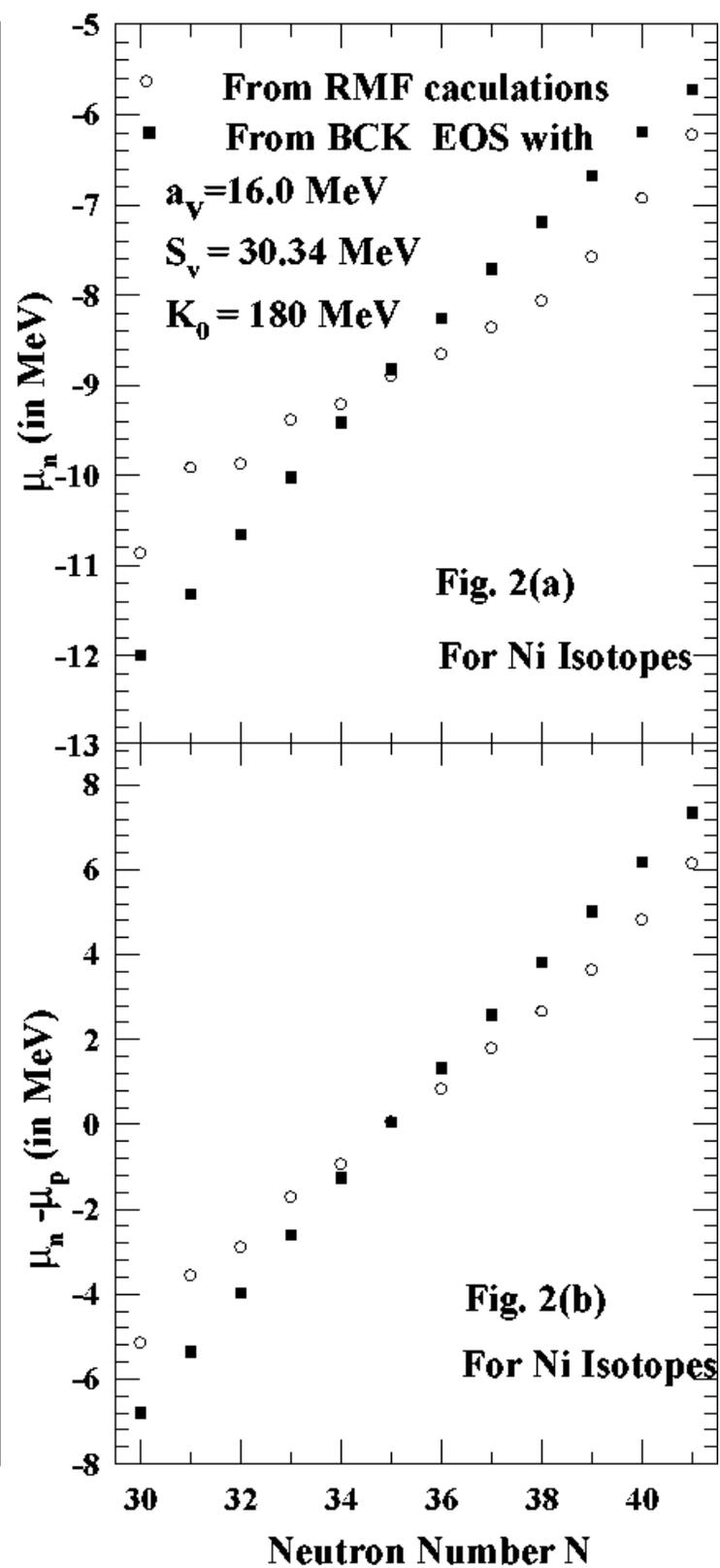
\end{document}